\documentclass[]{spie}  

\usepackage{amsmath,amsfonts,amssymb}
\usepackage{graphicx}
\usepackage[colorlinks=true, allcolors=blue]{hyperref}

\title{VLTI Unit Telescope coudé train vibration control upgrade for GRAVITY+}

\author[a]{Romain Laugier}
\author[b]{Julien Woillez}
\author[a]{Denis Defrère}
\author[c,i]{Benjamin Courtney-Barrer}
\author[a]{Muhammad Salman}
\author[b]{Babak Sedghi}
\author[c]{Roberto Abuter}
\author[a]{Azzurra Bigioli}
\author[d]{Maximilian Fabricius}
\author[d]{Frank Eisenhauer}
\author[b]{Frédéric Gonté}
\author[c]{Nicolas Schuhler}
\author[d]{Dieter Lutz}
\author[c]{Miguel Riquelme}
\author[c]{Pierre Bourget}
\author[a]{Philippe Neuville}
\author[e,f,b]{Sylvestre Lacour}
\author[g,h,e]{Mathias Nowak}
\affil[a]{Institute of Astronomy, KU Leuven, Celestijnenlaan 200D, 3001 Leuven, Belgium}
\affil[b]{ESO Garching, Germany}
\affil[c]{ESO Paranal, Chile}
\affil[d]{Max Plack Institut für Extraterrestrische Physik, Germany}
\affil[e]{Laboratoire d'Etudes Spatiales et d'Instrumentation en Astrophysique, Observatoire de Paris, France}
\affil[f]{Université PSL, Sorbonne Université, CNRS, France}
\affil[g]{Institute of Astronomy, University of Cambridge, United Kingdom}
\affil[h]{Kavli Institute for Cosmology, Cambridge, United Kingdom}
\affil[i]{Research School of Astronomy \& Astrophysics, Australian National University, ACT 2611, Australia}

\affil[d]{Université Côte d'Azur, Observatoire de la Côte d'Azur, CNRS, Laboratoire Lagrange, France}
\affil[e]{The University of Sydney, Australia}
\affil[f]{Univ. zu Köln, Germany}

\authorinfo{Further author information: (Send correspondence to R. Laugier)\\R. Laugier: E-mail: romain.laugier@kuleuven.be}

\begin{document} 
\maketitle

\begin{abstract}
Fringe stability and tracking are a determining aspect for the performance of current interferometric observations. While the theory predicts that the aperture of large telescopes such as the VLTI UT should yield smoothed-out piston perturbations that could be compensated using a slow fringe tracker running at a few tens of Hz, this is far from the current experimental reality. In practice, the optical path variations observed with the GRAVITY fringe tracker still contain high frequency components that limit the fringe-tracking exposure time and therefore its precision and limiting magnitude.
Most of these perturbations seem to come from mechanical vibrations in the train of mirrors, leading to the instrument, and in particular from the mirrors of the telescope. With this work, and as part of the GRAVITY+ efforts, accelerometers were added to all the mirrors of the coudé train, including the coming M8, to complement the existing instrumentation of M1, M2, and M3, and compensate in real-time the optical path using the main delay lines.
We show how the existing architecture, while optimal for the first mirrors, is not suitable for the vibration content found in the new mirrors, and we opt instead for narrow-band filters based on phase-locked-loop filters (PLL). Thanks to this architecture, we were able to reclaim up to 50nm of OPD RMS from vibrations peaks between 40 and 200Hz.
We also outline the avenues to push this approach further, through the upgrade of the deformable mirrors and the beam-compressor differential delay lines (BCDDL) as part of GRAVITY+, paving the way to obtaining better than 100nm RMS fringe tracking, even on faint targets.

\end{abstract}

\keywords{Interferometry, fringe-tracking, GRAVITY+, vibration control, VLTI}

\section{Introduction}\label{sec:intro}
    The GRAVITY instrument has demonstrated the potential of phase referenced optical interferometry for the study of planetary systems, active galactic nuclei and to probe the limits of general relativity. The ongoing GRAVITY+ upgrade aims to increase the sensitivity to enable observation of fainter targets, through improvements of the instrument and of the VLTI infrastructure. One of the current limitations is the capability for high precision fringe tracking and the capability to track on faint reference stars \cite{Lacour2019a, Nowak2024a}. Inherent limitations of fringe tracking are include the temporal frequency content of the phase errors. The correlation of spatial and temporal components of the the wavefront errors \cite{Conan1995} suggest that the use of adaptive-optics (AO) corrected larger telescopes such as the 8m unit telescopes of VLTI would lead to a more favorable power law of $-17/3$ in temporal frequency \cite{Courtney-Barrer2022}. The resulting coherence times should allow to increase the fringe-tracker exposure times and lower loop frequency to a few tens of Hz, increasing the achievable precision on bright targets, and enabling the use of fainter targets.

\begin{figure}
        \centering
        \includegraphics[width=0.9\textwidth]{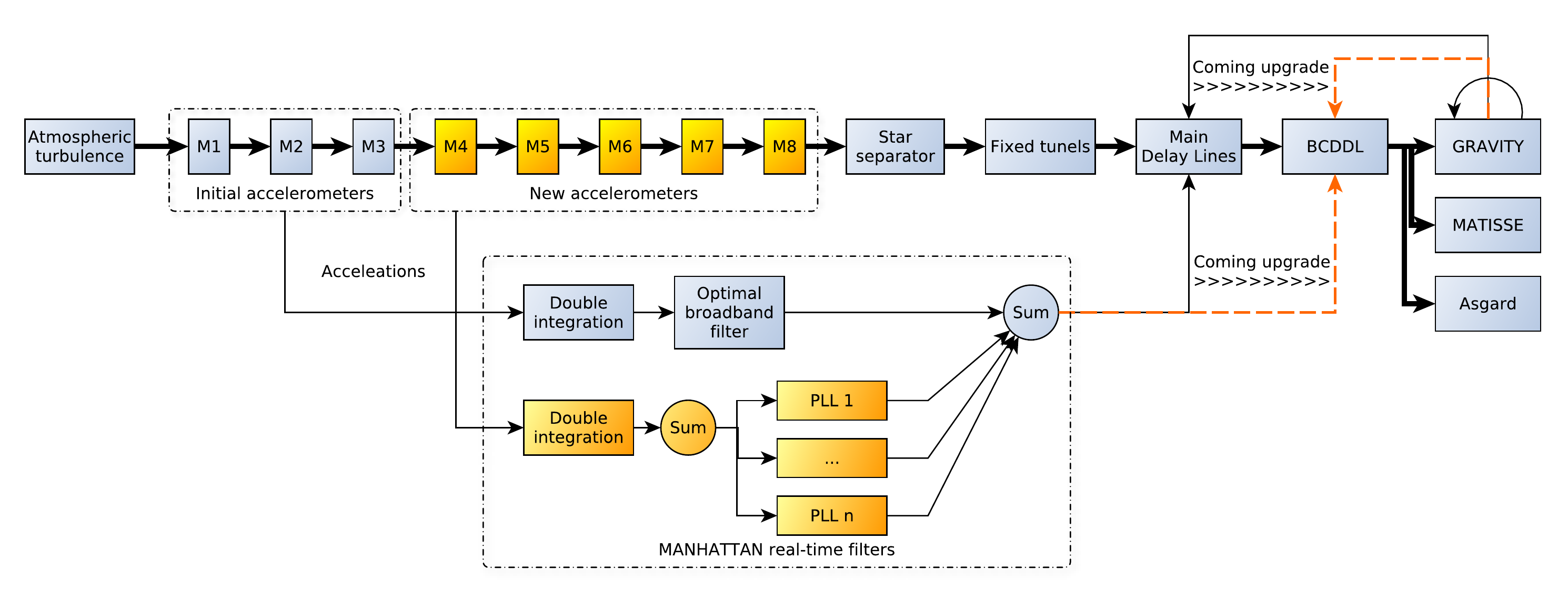}
        \caption{Schematic representation of the different aspects of the upgrade, showing in orange the new mirrors included and their use in the narrow-band PLL filters, as well as the ongoing transition to the BCDDL for actuation.}
        \label{fig:poster_schematic}
    \end{figure}

\begin{figure}
        \centering
        \includegraphics[width=0.5\textwidth]{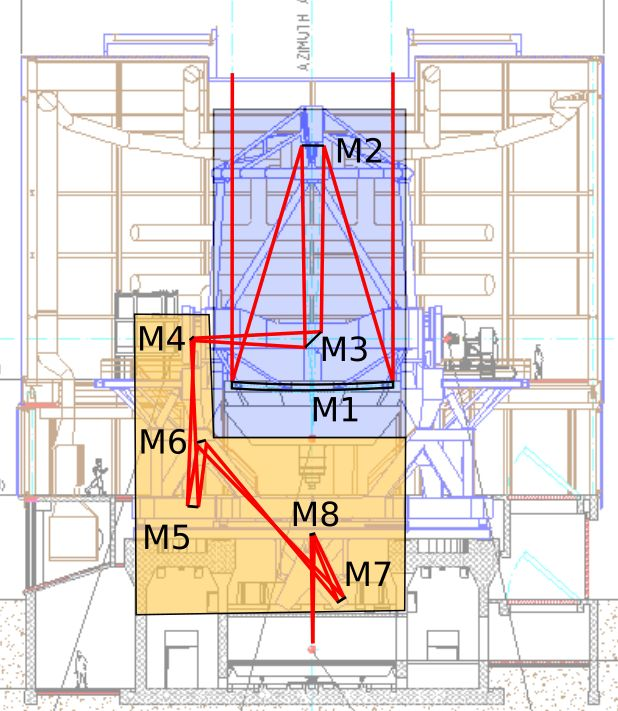}
        \caption{The schematic representation of the mirrors encompassed by the initial and upgraded version of MANHATTAN system}
        \label{fig:mirrors_old_new}
    \end{figure}

    While signs of this power low can be experienced with the 1.8m auxiliary telescopes (ATs), the power spectrum of optical path lengths with the unit telescopes (UTs) still contains a large amount of high-frequency content that limits the performance of all fringe trackers. These perturbations have been attributed to mechanical vibrations of the telescope mounts imprinted onto the beam by the mirrors of the coudé optics. The MANHATTAN system is a VLTI subsystem serving to mitigate the effect of these vibrations by measuring them with accelerometers mounted on the telescope mirrors, computing the resulting optical path errors and sending correction orders to the VLTI delay lines \cite{Haguenauer2008} in a feed-forward configuration. Prior to our upgrade, this included accelerometers on M1, M2 and M3 of each UT. As illustrated by Fig. \ref{fig:poster_schematic}, his upgrade includes instrumentation of all the mirrors of the coudé train (M4, M5, M6, M7 and M8\footnote{Accelerometer has not been installed on M8, which will become the deformable mirror of the new adaptive optics GPAO later in 2024, which will carry an accelerometer.} as shown in Fig. \ref{fig:mirrors_old_new}), improvement on the digital feed-forward filter, and a coming modification of the correcting actuator.

\section{Practical limitations}\label{sec:limitations}
\begin{figure}
        \centering
        \includegraphics[width=0.9\textwidth]{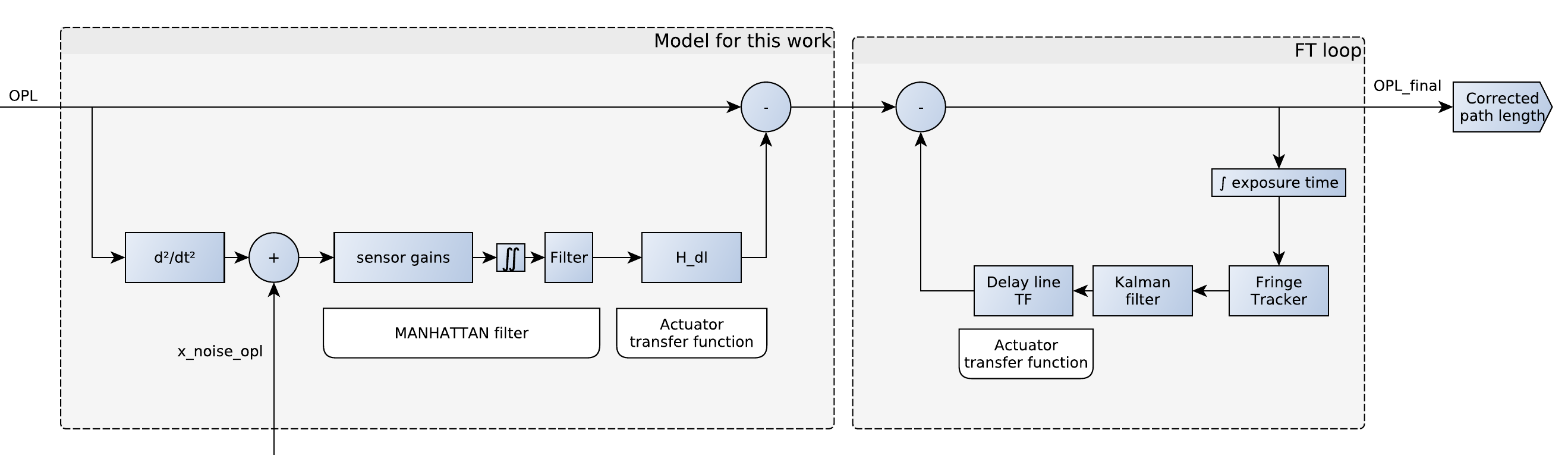}
        \caption{A schema-bloc representation of the MANHATTAN system, correcting the mechanical optical path errors in a feedforward configuration using the delay lines, based on measurements of the acceleration of the mirrors.}
        \label{fig:manhattan_concept}
    \end{figure}

\begin{figure}
        \centering
        \includegraphics[width=0.45\textwidth]{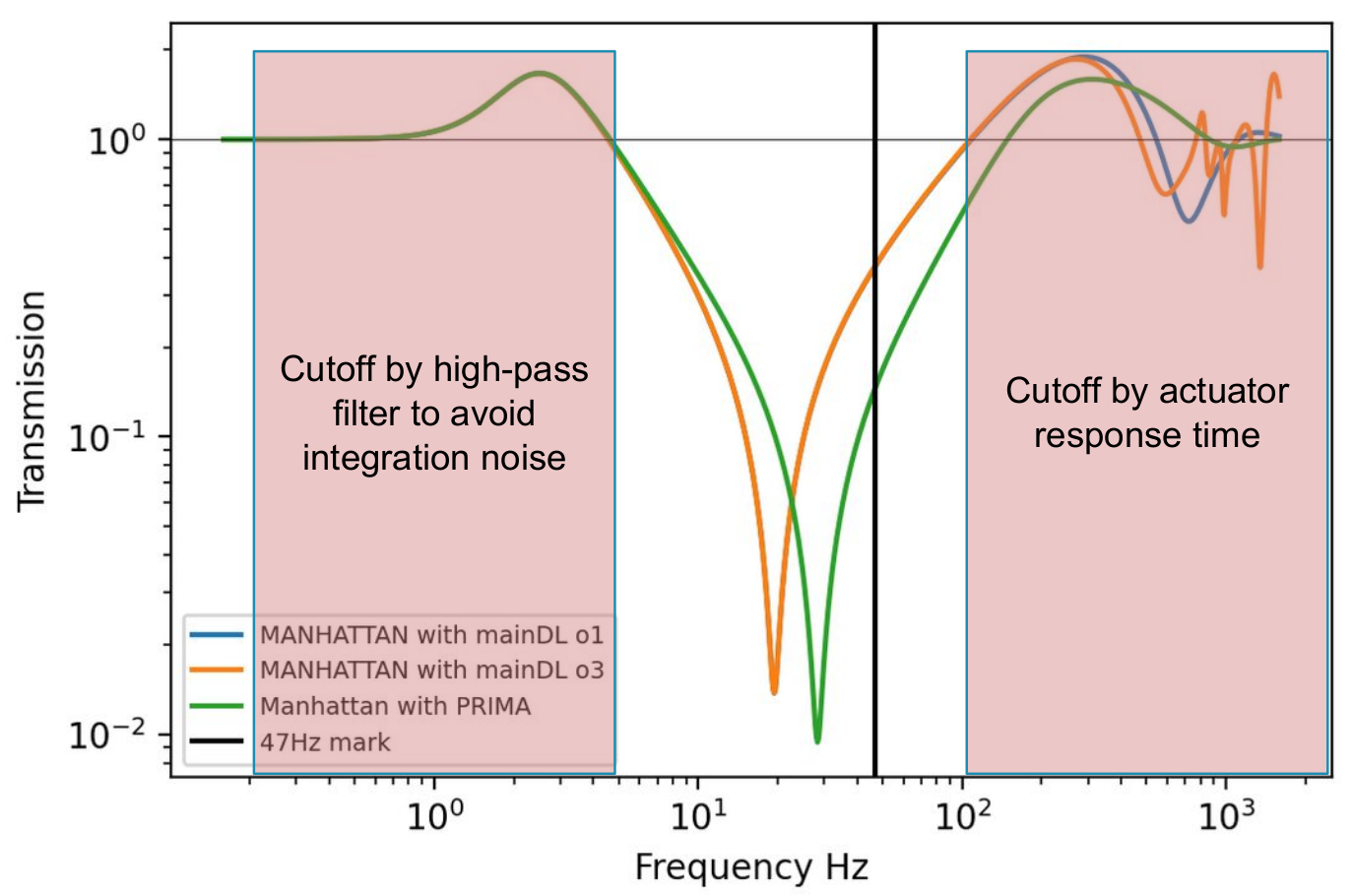}
        \includegraphics[width=0.45\textwidth]{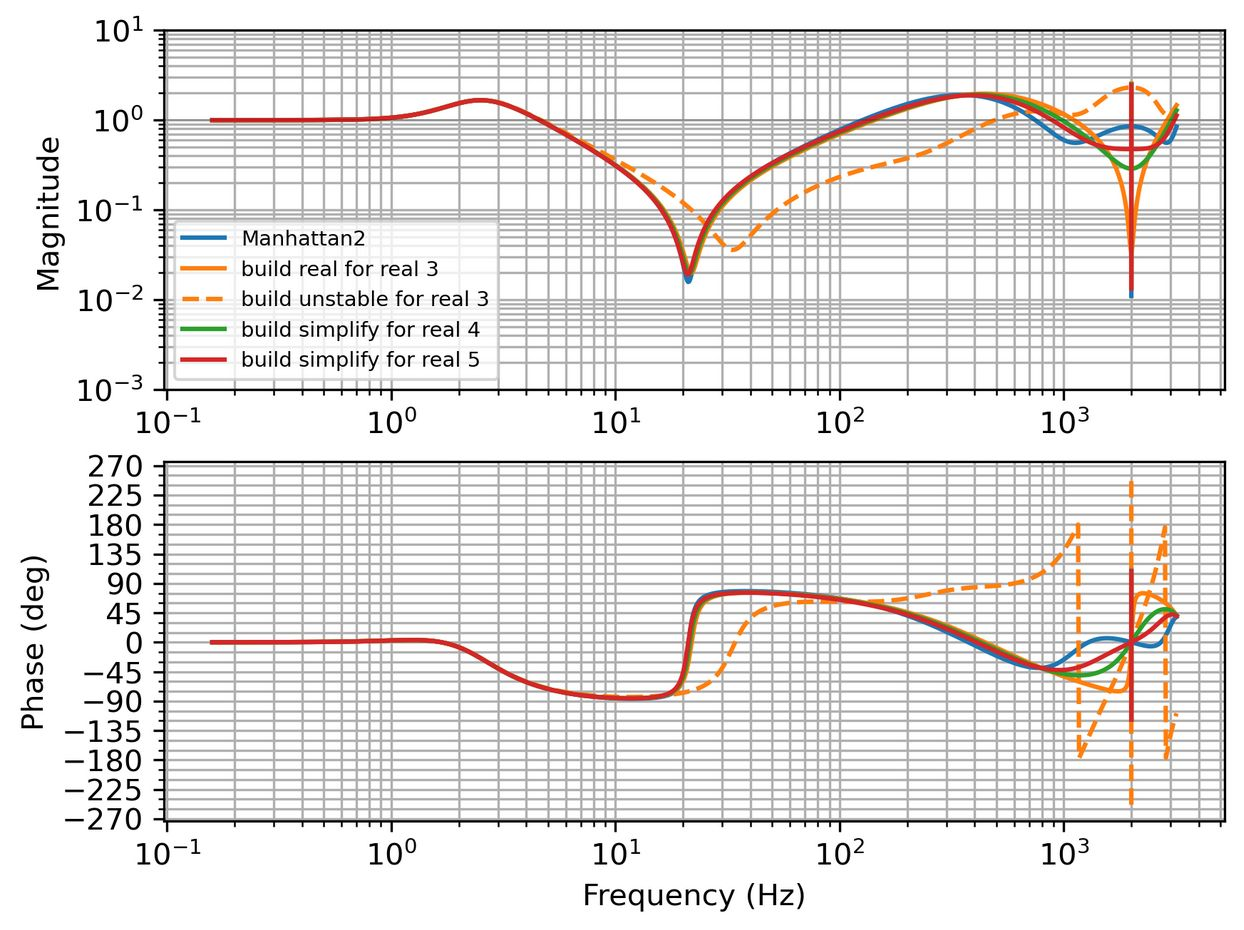}
        \caption{Representation of the transfer function of the manhattan feedforward correction of optical path. For this model, the Input is the mechanical optical path as offered by the passive system, and output is the output is the total optical path after correction by the active system. Left: showing the initial system and in red fundamental limitations by the sensor noise at low frequencies, and by the actuator response time at high frequencies. Right: detail on the effect of pure delay on the performance of a stable system. The dashed orange line is a theoretical system that would not be limited by causality or stability. After regularization, the system comes back to the initial performance.}
        
        \label{fig:limitations}
    \end{figure}

    As illustrated in Fig. \ref{fig:manhattan_concept}, the feedforward algorithm of MANHATTAN operates at 4kHz on the optical path of each beam, independently of the closed loop GRAVITY FT which optically measures and corrects the atmospheric perturbations and any leftovers from the the mechanical perturbations to the optical path. MANHATTAN is implemented on a real-time local control unit (LCU) using the tools for advanced control (TAC) library provided by ESO. Its sensors measure accelerations and retrieves a position measurement via double integration, before applying an ad-hoc filter and sending the correction instruction to the delay line actuator via the VLTI reflective memory network (RMN). This filter is designed to best negate the effect of the transfer function (TF) of the actuator. The MANHATTAN feedforwad filter is defined in OPL from its ground truth to its value as corrected by the actuator (left block of Fig. \ref{fig:manhattan_concept}). Its performance is the rejection of this TF as shown in Fig. \ref{fig:limitations}.
    
    \paragraph{Low frequency limits}
        Noise on the acceleration measurement is dependent on the sensor technology. With the current implementation the piezoelectric accelerometers the noise floor is considered white, and scales with the length of the cables (from their capacitance) \cite{Bigioli2022}. Due to the necessary double integration, this leads to a red spectrum, which is mitigated by a second order Butterworth high-pass filter which prevents the divergence of the system. The effect of this filter and its cutoff frequency set to 5 Hz limits the performance of MANHATTAN in the lower frequencies.

\begin{figure}
        \centering
        \includegraphics[width=0.45\textwidth]{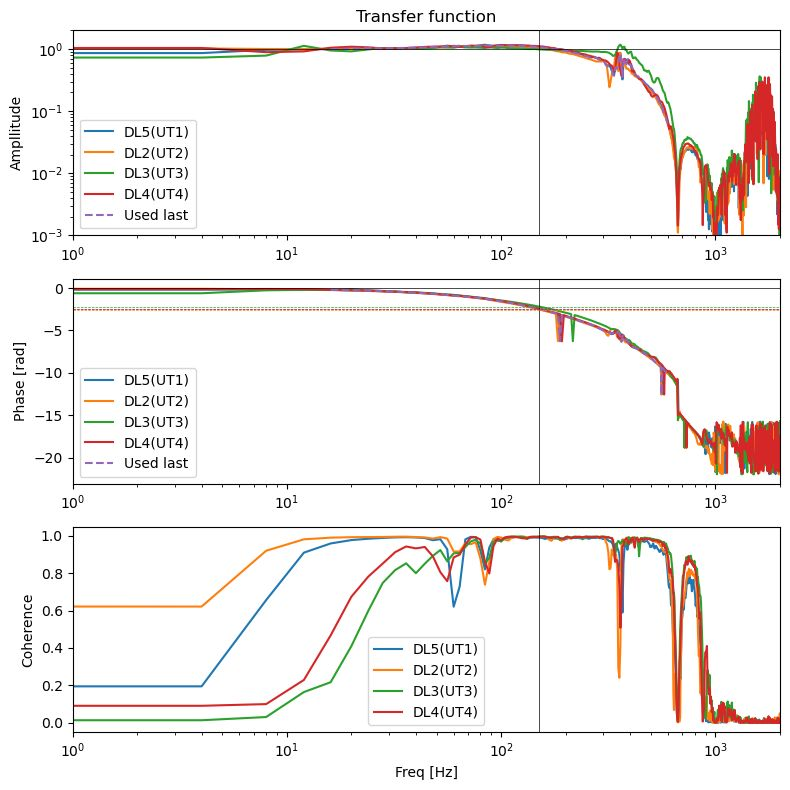}
        \caption{Measurement of the transfer function of the delay lines of VLTI through the MANHATTAN actuation path, as measured by their metrology system. While the amplitude response is almost flat up to 300Hz, the phase response is consistent with pure delay of 2.2 to 2.6 ms in addition to the actuators electro-mechanical response.}
        \label{fig:dl_tf}
    \end{figure}

    \paragraph{High frequency limits}
        To obtain ideal response in feedforward, the TF of the filter should be the inverse of the TF of the actuator. However, the existence of pure delay in this actuation (i.e. it is not minimum-phase) means that a direct inverse would lead to a non-causal transfer function, which cannot be implemented. The accepted approach \cite{Huang, Maggio2020}, is to use an orthogonal decomposition of this TF into a minimum-phase system and an all-pass filter which contains the pure delay of the system, including the computing time of the LCU and the pure delay of the actuator response. Using MANHATTAN, to send white noise through this command path, we used the internal metrology of the DL to measure their TF as shown in Fig. \ref{fig:dl_tf} shows pure delay of 2.2 to 2.6 ms in addition to the actuators electro-mechanical response. As illustrated by Fig. \ref{fig:limitations}, this is dominant in defining the high frequency limits of the MANHATTAN performance as multiple attempts to improve upon this filter were brought-back by this regularization to a behavior very close the that of the initial implementation.

    As a consequence, the MANHATTAN system filter is a band-cut filter between over a range between 5 Hz and 100 Hz, for which the performance is limited by the fundamental effects of practical limitation. Working at lower frequencies would require an upgrade in accelerometers technology, and improving rejection at higher frequencies for broadband control would require an upgrade of the actuators which is discussed in Sect. \ref{sec:actuator}. For the correction of the high-frequency signals measured in the new accelerometers, a workaround is described in Sect. \ref{sec:narrowband} using narrow-band filters.

\section{Narrow-band filters}\label{sec:narrowband}
    The OPL variations can be measured optically from the optical path differences (OPDs) measured by the GRAVITY FT at 980Hz projected back onto the four beams and adding the optical path applied by its own actuators. This result is called the pseudo-open-loop (POL) measurement, and includes the contributions of the atmosphere and of the mechanical vibrations. The power spectral density (PSD) is computed and the square root of the reverse-cumulative PSD helps identify the peaks of largest contribution.
    
\begin{figure}
        \centering
        \includegraphics[width=0.45\textwidth]{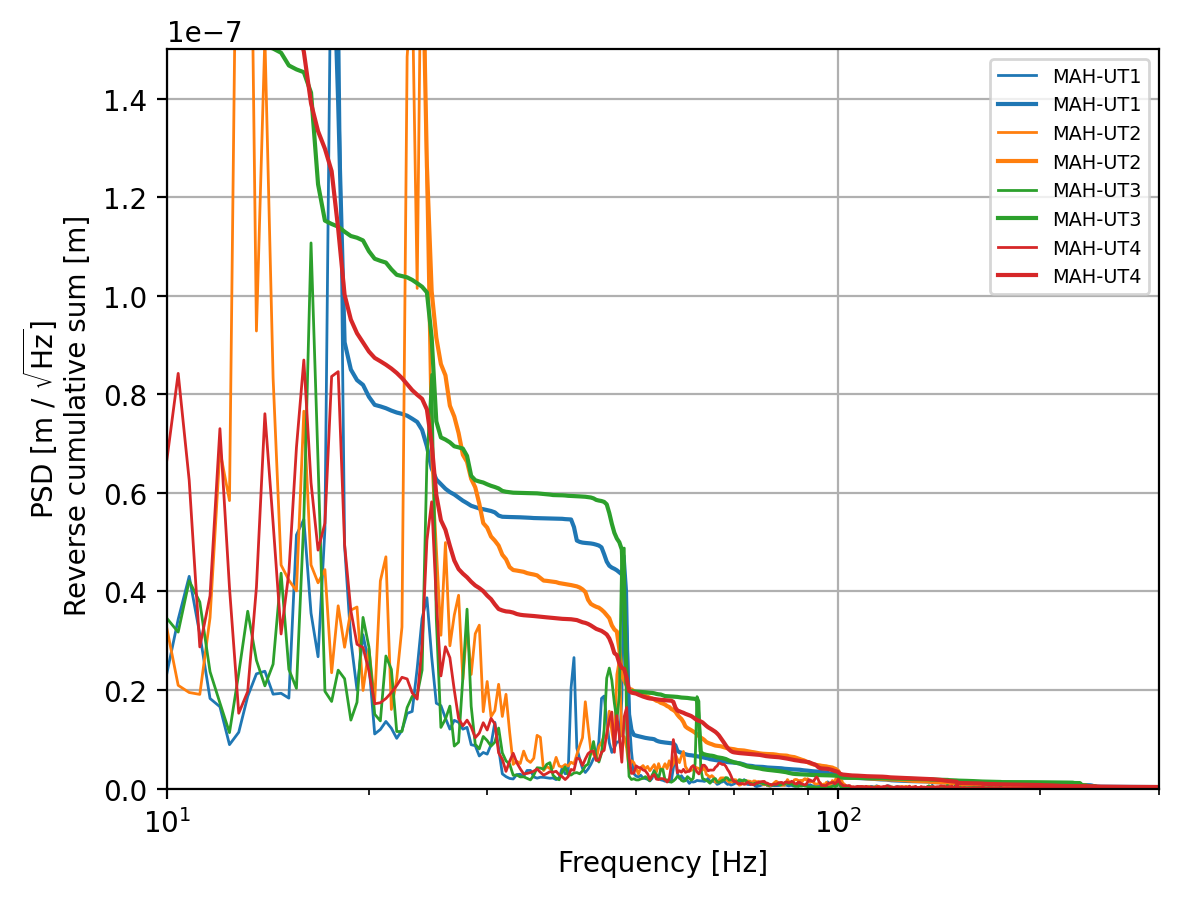}
        \includegraphics[width=0.45\textwidth]{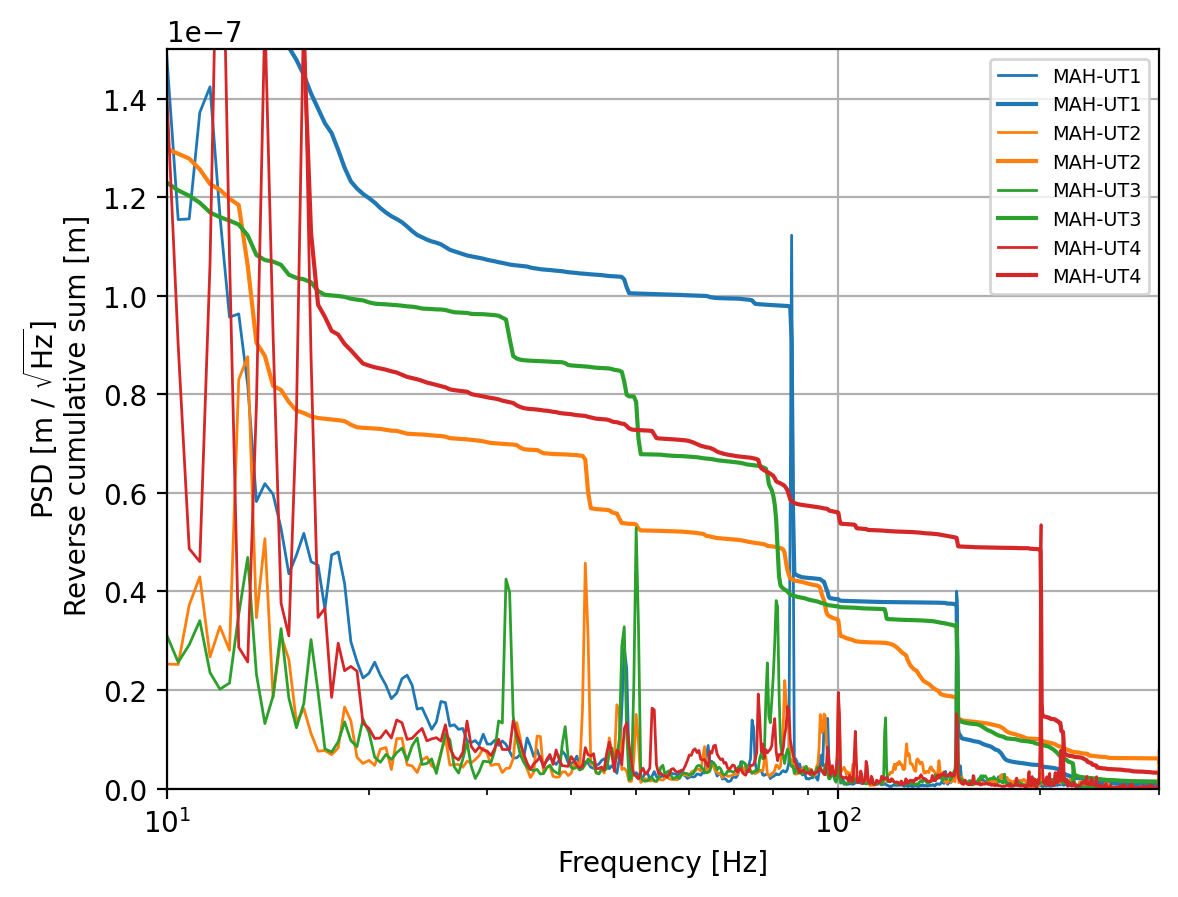}
        \caption{Comparison of the vibration content in the pre-existing MANHATTAN sensors M1 to M3 on the left, and on the newly instrumented mirrors M4 to M7 on the right. While the vibration content on the first three mirrors appears below 50-70Hz, the mirrors of the coudé train exhibit strong vibrations around 80, and at 150 and 200Hz in narrow peaks.}
        \label{fig:acc_opl_new_mirrors}
    \end{figure}

    An analysis of the signal provided by the new and pre-existing sensors was carried out as illustrated by Fig. \ref{fig:acc_opl_new_mirrors}. While the first three mirrors show mostly vibrations between 10 and 50Hz, the new mirrors exhibit strong narrow-band peaks around 80 and at 150 and 200Hz. Furthermore, it was found that the motion of these mirrors showed a form of coherence as they belong to a common elastic structure (the azimuthal fork) which vibrates following a number of harmonic modes.
    
    Due to the fundamental limitations mentioned in the previous section, implementing such corrections in a broadband filter was not feasible. However, narrow-band filters can circumvent this type of limitations thanks to the periodicity of the narrow-band components which allows us to mimic an advance in phase.

\begin{figure}
        \centering
        \includegraphics[width=0.75\textwidth]{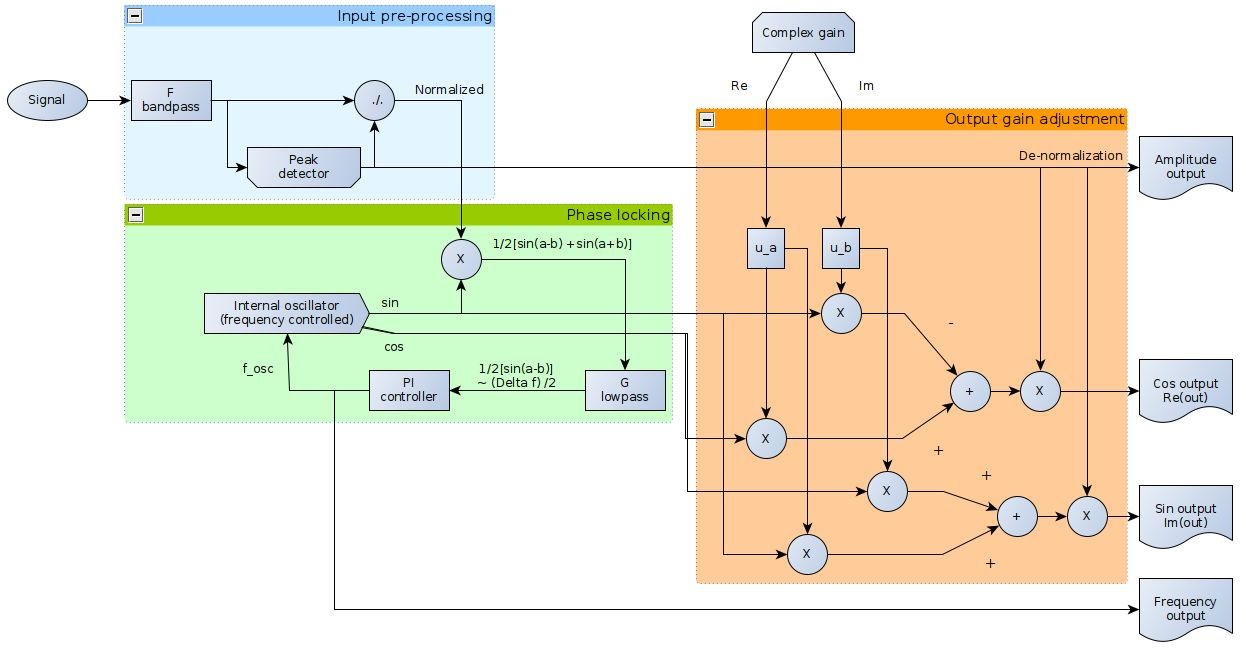}
        \caption{Schematic of the custom PLL filtered of the TAC library. The filter was modified to include a tracking of amplitude in addition to phase, using a peak-detector algorithm.}
        \label{fig:pll_schematic}
    \end{figure}

    Here we implement a phase-locked loop (PLL) filter from TAC modified as shown in Fig. \ref{fig:pll_schematic} to track both the phase and amplitude of the signal of interest. After a second-order Butterworth bandpass filter, the input signal is mixed with that of an internal synthetic oscillator which gives the signal $m$:
    \begin{equation}
        m = \frac{1}{2} \Big(\sin(a - b) + \sin(a+b) \Big)
    \end{equation}
    where $a$ and $b$ are the phase term of the phase term of the input signal and the internal oscillator. The term in $\sin(a+b)$ is discarded using low-pass filter and the term in $\sin(a-b)$ is used in the linear approximation in a proportional-integral (PI) servo loop controlling the frequency of the internal oscillator to bring this phase difference to zero.

    The output of this filter offers both the real and imaginary part of this output signal, which allows us to add arbitrary phase and amplitude corrections to account for the TF of the actuator, including the necessary positive phase offset. In the most recent upgrade, an adaptive compensation is even added to correct the effect of the input bandpass filter. 

\begin{figure}
        \centering
        \includegraphics[width=0.8\textwidth]{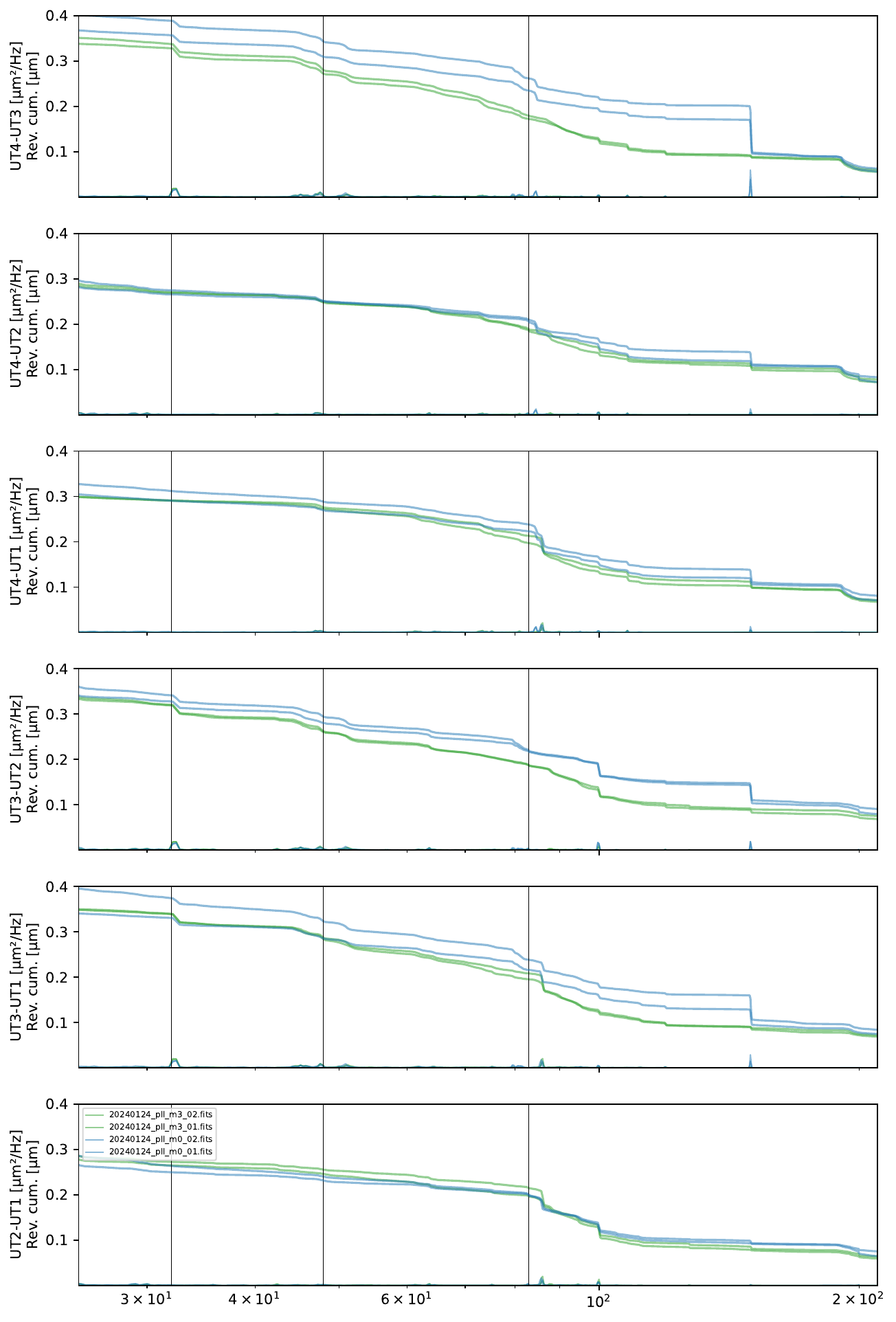}
        \caption{The GRAVITY FT POL signal for all of the baselines during obtained on a bright calibration star. The PLL filters were experiment followed a sequence on-off-on-off to limit account for the variability of the observing conditions. The plot highlight the square root of the reversec-cumulative PSD which gives the RMS error cumulated from high to loo frequencies in µm (off in blue and on in green). }
        \label{fig:gravity_pol}
    \end{figure}

    The effectiveness of these filters was verified on-sky using the gravity POL signal showed in Fig. \ref{fig:gravity_pol}. This approach was very effective in correcting the vibrations peaks found at 200Hz and 150Hz for which an improvement of around 100 nm RMS was observed on some of the baselines. Using mechanical information of MANHATTAN to correct for such high-frequency content is crucial as it is very difficult to measure them with the FT on fainter targets, where it must run with longer integration times.

\section{Actuator upgrade}\label{sec:actuator}

    As shown in Sect. \ref{sec:limitations}, the improvement of the correction by the broadband filter is limited by the actuator delay. Thanks to the GRAVITY+ upgrade a new set of differential delay lines (BCDDLs) is being installed in the VLTI recombination lab, and their response time is shorter than that of the main DL. Due to the much smaller pure delay term, this actuator is expected to help tackle the high-frequency limitations for the broadband filter (see presentation [13095-10]).

\begin{figure}
        \centering
        \includegraphics[width=0.9\textwidth]{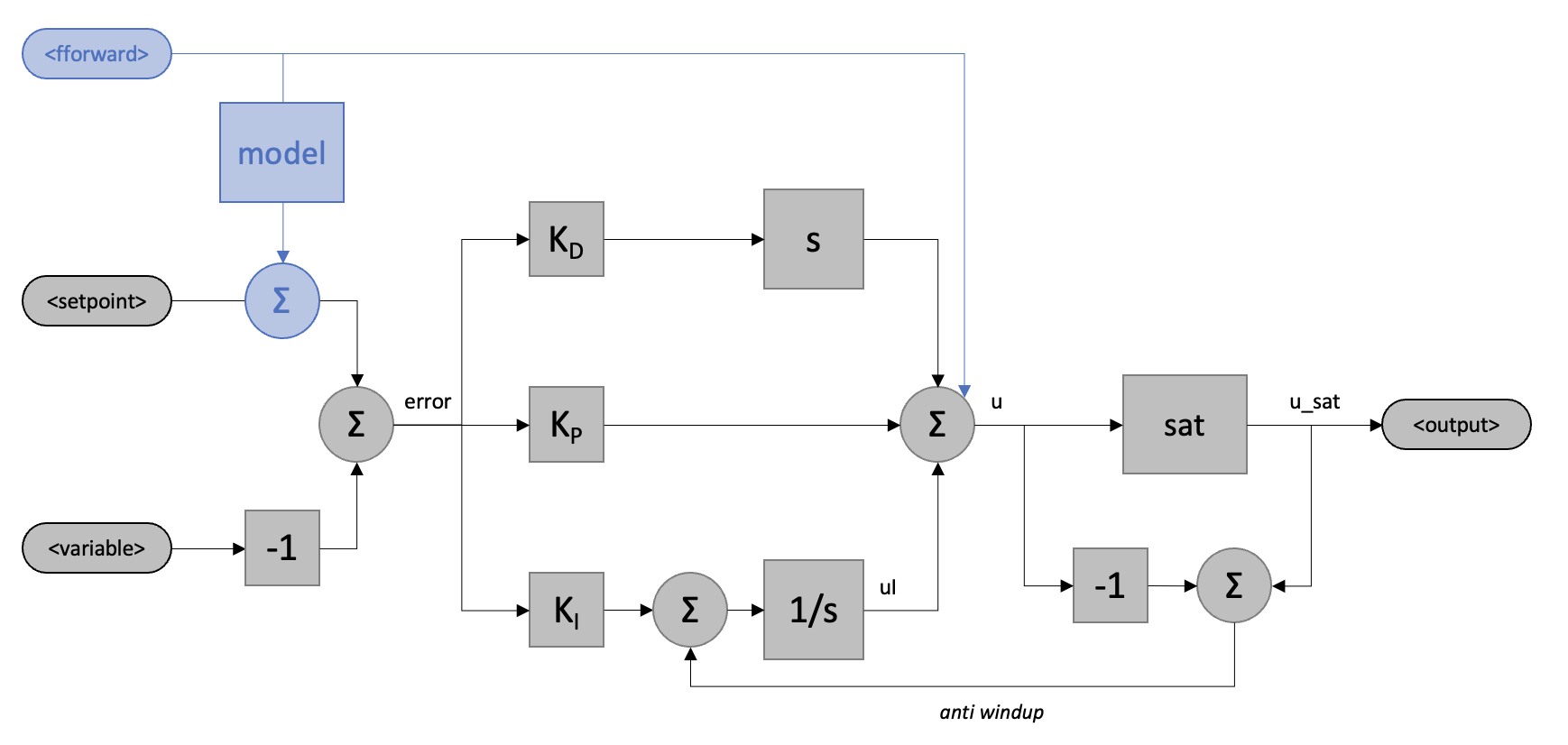}
        \caption{Schema-bloc representation of the BCDDL control. The blocks in grey are involved in the normal PID closed loop system. The blocks in blue are used to bypass the servo loop and send commands directly to the piezo actuator. The block \texttt{model} contains the transfer function of this bypass piezo system, as measured when the PID is deactivated, allowing the servo controller to ignore the commands sent via the bypass command.}
        \label{fig:bcddl_control}
    \end{figure}

    The control system for the BCDDL reuse some of the control hardware of the PRIMA delay lines and uses a proportional integral derivative (PID) controller that takes a setpoint from the reflective memory network system and can be controlled like the main delay lines. In an attempt to further increase the bandpass, an additional control path was added which bypasses the main PID controller, allowing manhattan to control the piezo actuator directly as illustrated by Fig. \ref{fig:bcddl_control}.
    
    The transfer function, as seen by manhattan through the RMN was measured by injecting white noise. For the initial measurement of the bypass control, the piezo response was measured while turning off the PID loop, so as to avoid antagonistic behavior. In operation, the TF measured here will implemented in the block labeled \texttt{model} to avoid allow parallel usage of both modes.

    \begin{figure}
        \centering
        \includegraphics[width=0.75\textwidth]{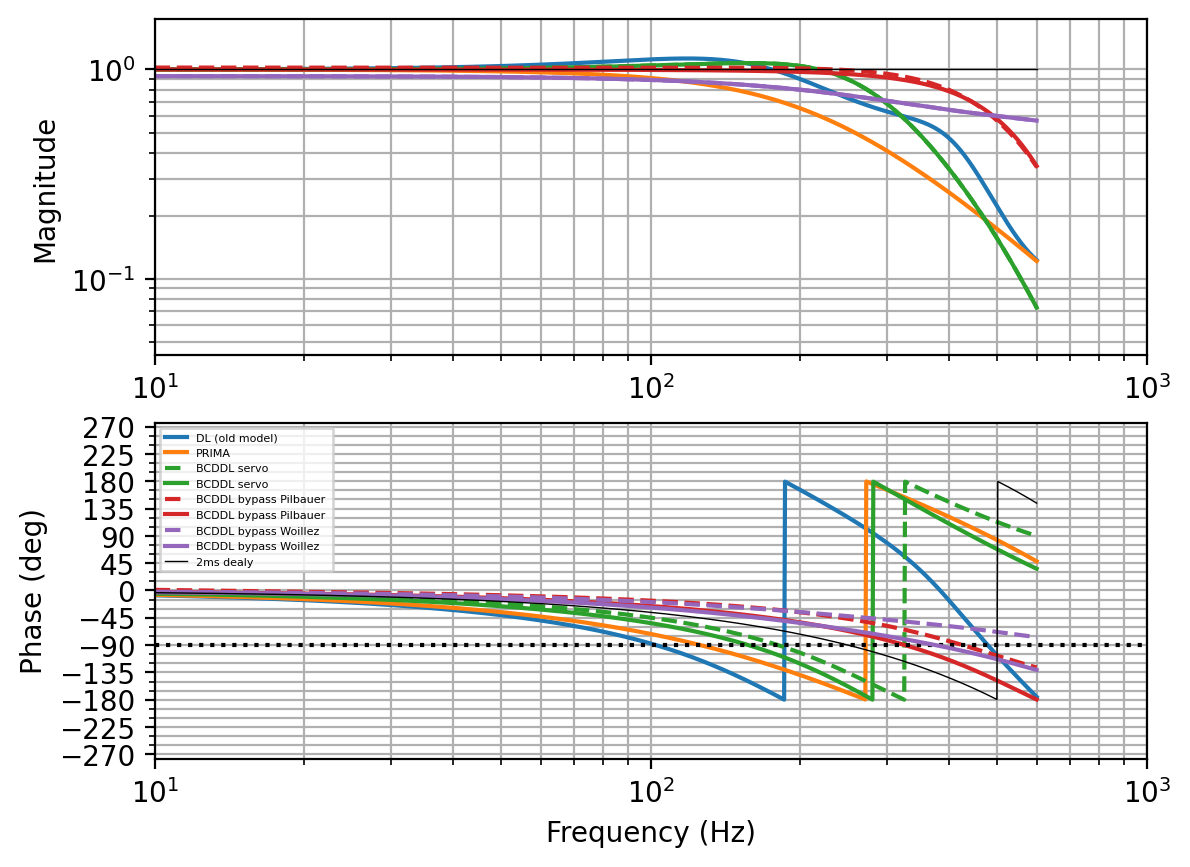}
        \caption{Comparison of the TF of different piston actuators. Dashed line indicate the response before considering the 250µs lag of the digital computation by the LCU. The main delay lines are the slowest. The BCDDL in bypass mode appear to be the fastest, as indicated by their phase response, however it suffers from a lowered gain due to its open-loop configuration.}
        \label{fig:compared_actuators}
    \end{figure}

\begin{figure}
        \centering
        \includegraphics{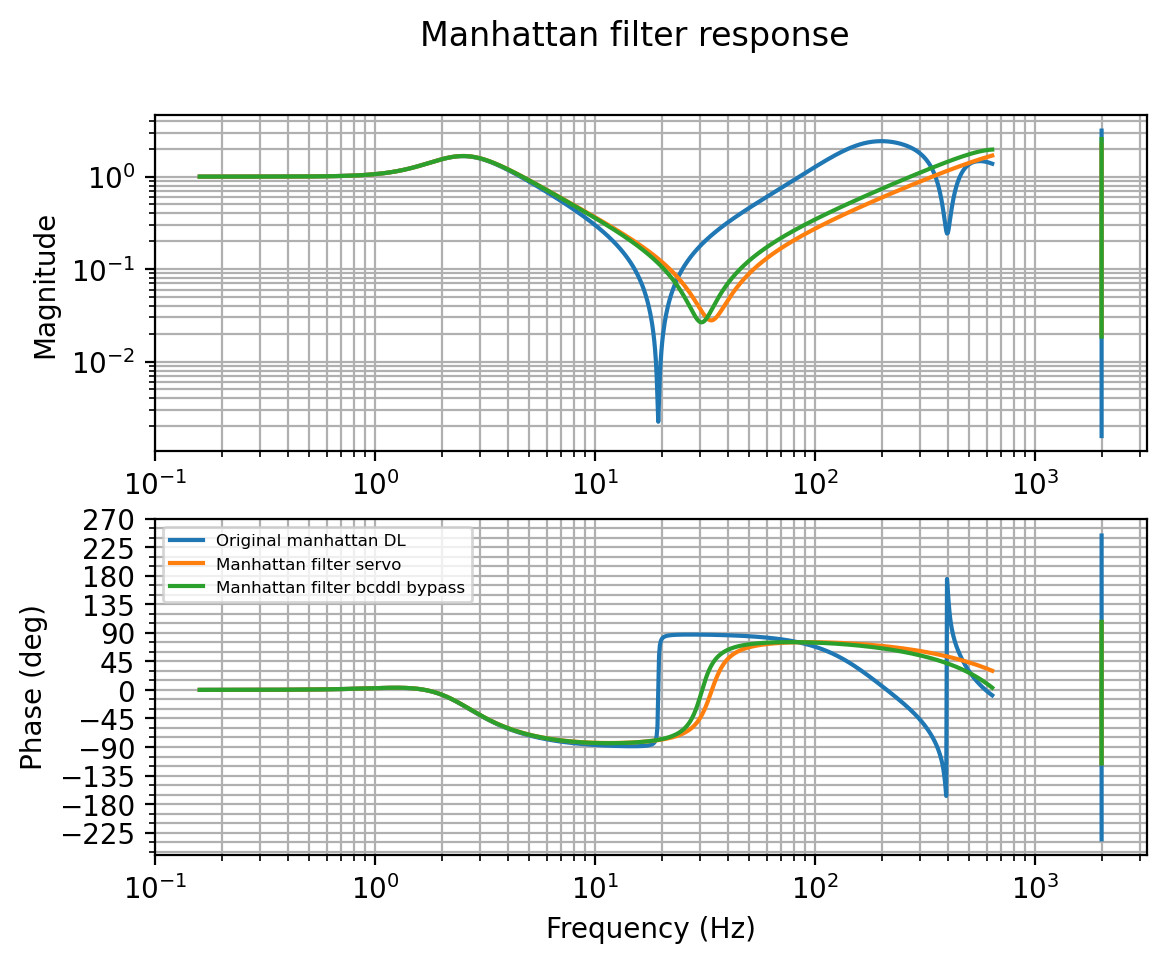}
        \caption{Comparison of the expected MANHATTAN performance with the different available actuation modes}
        \label{fig:compared_perf}
    \end{figure}

    Using again the approach accepted approach, we compute an optimal stable inverse of the actuator's transfer function for both servo and bypass mode of BCDDL. This produces order 10 and 11 filters to be implemented for the broadband MANHATTAN controlling mirrors M1-3. The rejection performance of the system is compared in Fig. \ref{fig:compared_perf}. We conclude that the control with the BCDDL offers a significant improvement of the rejection cutoff from 90Hz to about 300Hz, leading to highly improved rejection between 30 to 100Hz where significant vibration contributions remain for these mirrors. However, it appears the that despite a faster response, the bypass mode does not lead to a better feedforward rejection in system.

\section{Conclusion and perspectives}\label{sec:conclusion}

    The correction of the effects of mirror vibrations on optical path in feedforward is limited in the low-frequencies by the noise floor of the sensors which imposes a high-pass filter cutting the response at around 3-5Hz. In the high-frequencies, a fundamental limit arises from the response time of the system, currently dominated by pure delay in the main delay lines.

    The upgrade of the MANHATTAN system described here is focused on the incorporation of new accelerometers covering the whole coudé train, and the correction of the higher frequency optical path errors, which would not be seen by a longer exposure time fringe tracker. We circumvent the limitations by using narrow-band filters targeting known vibration peaks over which a phase advance can be produced. This was successful on the strong 200 and 150Hz vibrations.
    
    To overcome the high frequency limitation in broadband correction, we plan to use the new and faster BCDDL actuators for this control, extending the performance of the broadband correction up to some of the most prominent peaks, and implementation is still in progress, and should be commissioned later in 2024.

\begin{figure}
        \centering
        \includegraphics[width=0.8\textwidth]{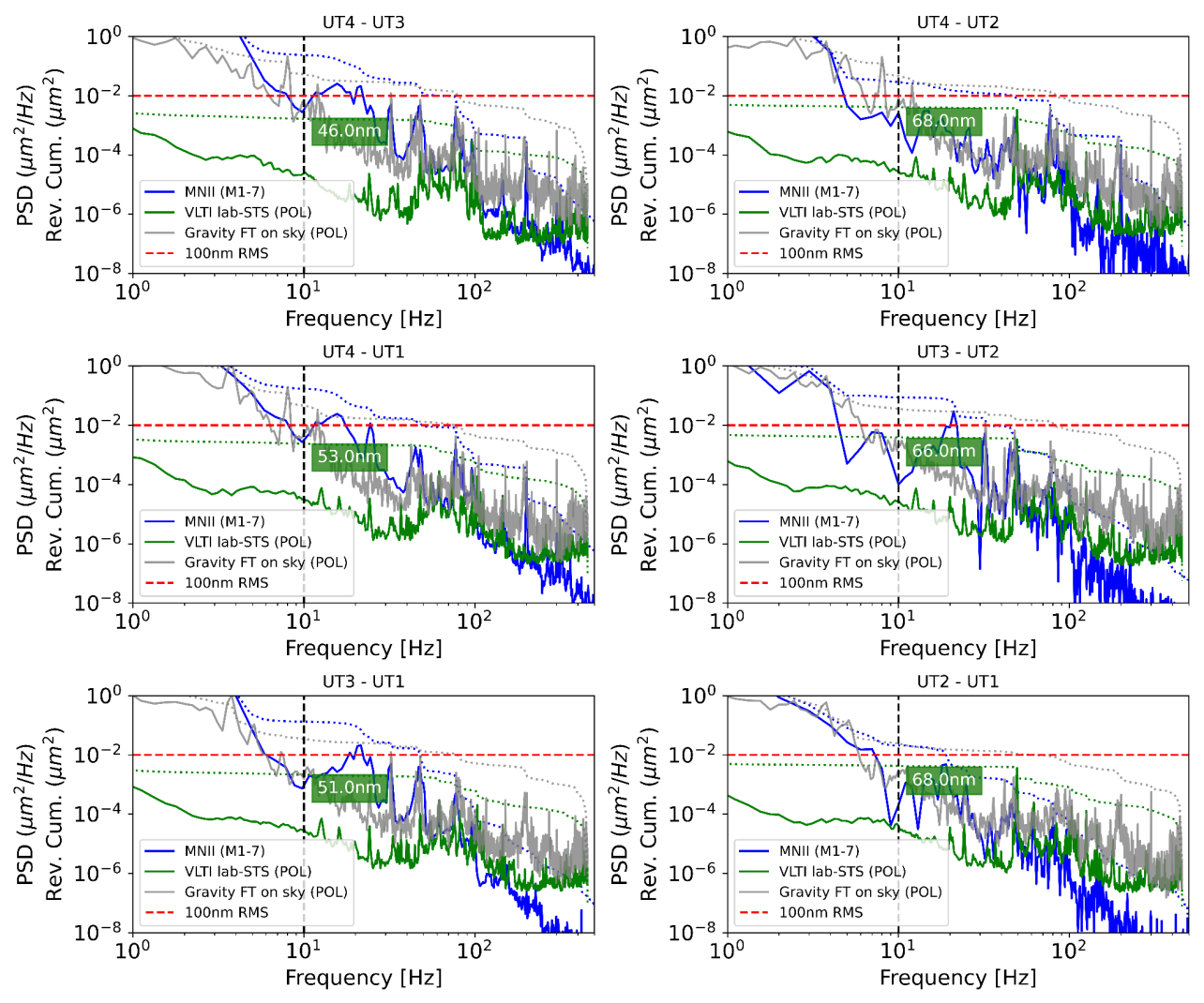}
        \caption{Optical path difference in green as measured in daytime by the autocollimation of the Marcel source sent throught the tunels to the star separator (STS) and back to the GFT thanks to a retroreflector, as compared to the MANAHATTAN measurement in blue and the on-sky pseudo open-loop measurement by GFT in gray. This shows a contribution of about 46-68 nm from tunnel optics, in particular around 50-90 Hz. }
        \label{fig:autocol}
    \end{figure}

    The current network of sensors is limited to M7 with a coming extension to M8 with the implementation of GPAO \cite{Eisehauer2020}. It is blind to the vibrations coming from the rest of the optical train, through the VLTI tunnel and the DL. The vibration content in the tunnels has been measured by an autocollimation measurement using Marcel to send light back to the star separator (STS) onto a retroreflector and back to the GFT. The result of this measurements, shown in Fig. \ref{fig:autocol} shows a total of around 45-65 nm RMS errors, mostly coming from a number of peaks between 50 and 90Hz, which could explain our inability to correct vibrations in this range with the current system.
    
    Among the high frequency components, a broad vibration peak is detected at 190Hz which is not seen by the accelerometers. As this corresponds to resonant frequencies of the MACAO deformable mirror, it cannot be measured by MANHATTAN. However it should disappear with the replacement of deformable mirror later this year, offering even better stability at the high frequencies.

    Further improvement in the lower frequencies would require a change of accelerometer technology,

\acknowledgments 
This work has received funding from the Research Foundation -  Flanders (FWO) under the grant number 1234224N. SCIFY has received funding from the European Research Council (ERC) under the European Union's Horizon 2020 research and innovation program (grant agreement CoG - 866070).

\bibliography{manhattan} 

\begin{thebibliography}{1}

\bibitem{Lacour2019a}
Lacour, S., Dembet, R., Abuter, R., F{\'e}dou, P., Perrin, G., {Choquet}, Pfuhl, O., Eisenhauer, F., Woillez, J., Cassaing, F., Wieprecht, E., Ott, T., Wiezorrek, E., Tristram, K.~R., Wolff, B., Ram{\'i}rez, A., Haubois, X., Perraut, K., Straubmeier, C., Brandner, W., and Amorim, A., ``The {{GRAVITY}} fringe tracker,'' {\em Astronomy and Astrophysics}~{\bf 624} (2019).

\bibitem{Nowak2024a}
Nowak, M., Lacour, S., Abuter, R., Woillez, J., Dembet, R., Bordoni, M.~S., Bourdarot, G., {Courtney-Barrer}, B., Defr{\`e}re, D., Drescher, A., Eisenhauer, F., Fabricius, M., Feuchtgruber, H., Frahm, R., Garcia, P., Gillessen, S., Gopinath, V., Graf, J., Hoenig, S., Kreidberg, L., Laugier, R., Bouquin, J. B.~L., Lutz, D., Mang, F., Millour, F., More, N., Moruj{\~a}o, N., Ott, T., Paumard, T., Perrin, G., Rau, C., Ribeiro, D.~C., Shangguan, J., Shimizu, T., Soulez, F., Straubmeier, C., Widmann, F., and Wolff, B., ``Upgrading the {{GRAVITY}} fringe tracker for {{GRAVITY}}+: {{Tracking}} the white light fringe in the non-observable {{Optical Path Length}} state-space,'' (Feb. 2024).

\bibitem{Conan1995}
Conan, J.-M., Rousset, G., and Madec, P.-Y., ``Wave-front temporal spectra in high-resolution imaging through turbulence,'' {\em Journal of the Optical Society of America A}~{\bf 12}(7),  1559 (1995).

\bibitem{Courtney-Barrer2022}
{Courtney-Barrer}, B., Woillez, J., Laugier, R., Bigioli, A., Schuhler, N., Guajardo, P., Lizana, V., Behara, N., Eisenhauer, F., Ireland, M., Haubois, X., and Defr{\`e}re, D., ``Towards a better understanding of {{OPD}} limitations for higher sensitivity and contrast at the {{VLTI}},'' in [{\em Optical and {{Infrared Interferometry}} and {{Imaging VIII}}}{\nolinebreak\hspace{0.1em}]},   {\bf 12183},  794--807, SPIE (Aug. 2022).

\bibitem{Haguenauer2008}
Haguenauer, P., Abuter, R., Alonso, J., Argomedo, J., Bauvir, B., Blanchard, G., Bonnet, H., Brillant, S., Derie, F., Delplancke, F., Di, N., Dupuy, C., Durand, Y., Gitton, P., Glindemann, A., Guniat, S., Guisard, S., Haddad, N., Hummel, C., Jesuran, N., Kaufer, A., Koehler, B., Le, B., L{\'e}v{\^e}que, S., Lidman, C., Mardones, P., M{\'e}nardi, S., Ramirez, A., Rantakyr{\"o}, F., Richichi, A., Rivinius, T., Sahlmann, J., Belle, G.~V., Wallander, A., Wehner, S., and Wittkowski, M., ``The {{Very Large Telescope Interferometer}} : An update,''  (October 2008) (2008).

\bibitem{Bigioli2022}
Bigioli, A., {Courtney-Barrer}, B., Abuter, R., Eisenhauer, F., Gonte, F., Laugier, R., Raskin, G., Riquelme, M., Salman, M., Schuhler, N., Woillez, J., and Defr{\`e}re, D., ``Measuring and compensating vibrations at the {{VLTI}}: {{MANHATTAN-II}} self-intrinsic noise and hardware extension,'' in [{\em Optical and {{Infrared Interferometry}} and {{Imaging VIII}}}{\nolinebreak\hspace{0.1em}]},  M{\'e}rand, A., Sallum, S., and {Sanchez-Bermudez}, J., eds.,  73, SPIE, Montr{\'e}al, Canada (Aug. 2022).

\bibitem{Huang}
Huang, D.-F., ``The {{Optimal Design}} of {{Stable Inverse Transfer Function}},''

\bibitem{Maggio2020}
Maggio, M., ``Feedforward {{Design}} - {{Real-Time Systems}}, {{Lecture}} 10,'' (2020).

\end{thebibliography}
\bibliographystyle{spiebib} 

\end{document}